\documentclass{article}
\usepackage{graphicx} % Required for inserting images
\usepackage{amsfonts}
\usepackage{amsmath}
\newtheorem{theorem}{Theorem}
\newtheorem{lemma}{Lemma}
\newtheorem{definition}{Definition}
\newtheorem{remark}{Remark}
\newtheorem{construction}{Construction}
\newtheorem{problem}{Problem}

\DeclareMathOperator\supp{supp}

\providecommand{\keywords}[1]
{
  \small	
  \textbf{\textit{Keywords---}} #1
}

\newtheorem{cor}{Corollary}

\title{Deformations and $q$-convolutions.\\ Old and new results}
\author{Marek Bo\.{z}ejko$^1$ \and Wojciech Bo\.{z}ejko$^2$\\[5mm]
$^1$Mathematical Institute\\
University of Wrocław\\
Pl. Grunwaldzki 2/4, 50-384 Wrocław, Poland\\
{\tt marek.bozejko@uwr.edu.pl}\\[2mm]
$^2$Department of Control Systems and Mechatronics\\
Wroc\l aw University of Science and Technology\\
Wyb. Wyspiańskiego 27, 50-370 Wrocław, Poland\\
{\tt wojciech.bozejko@pwr.edu.pl}
}

\date{\it \bf The paper is dedicated to Professor Jan Stochel on the occasion of his 70$^{th}$ birthday.}

\begin{document}

\maketitle

\begin{abstract}
This paper is the survey of some of our results related to $q$-deformations of the Fock spaces and related to $q$-convolutions for probability measures on the real line $\mathbb{R}$.
The main idea is done by the combinatorics of moments of the measures and related $q$-cumulants of different types.

The main and interesting $q$-convolutions are related to classical continuous (discrete) $q$-Hermite polynomial. Among them are classical ($q=1$) convolutions, the case $q=0$, gives the free and Boolean relations, and the new class of $q$-analogue of classical convolutions done by Carnovole, Koornwinder, Biane, Anshelovich, and Kula.

The paper contains many questions and problems related to the positivity of that class of $q$-convolutions. The main result is the construction of Brownian motion related to $q$-Discrete Hermite polynomial of type I.
\end{abstract}


\section{Introduction}

The plan of our note is the following:
\begin{enumerate}
    \item $q$-CCR(CAR) relations for $|q|>1$, and q-continuous Hermite polynomials.
    \item Combinatorial results on 2-partitions of $\{1,2,\ldots,2n\}$ -- $P_2(2n)$.
    \item q-discrete Hermite polynomials of type I, II.
    \item q-analogue of classical convolutions of Carnovale and Koornwinder \cite{CK} for $0 \leq q \leq 1$, ($q=0$, Boolean convolution, $q=1$ classical convolution).
    \item Braided Hopf algebras of Kempf and Majid.
    \item The construction of $q$-Discrete Fock space $\mathcal{F}_q^{disc}(\mathcal{H})$ and $q$-Discrete Brownian motions corresponding to $q$-Discrete Hermite polynomials of type I ($0 \leq q \leq 1$).
    \item Matrix version of Khintchine inequalities.
\end{enumerate}

\section{$q$-CCR(CAR) for $|q|>1$, and q-continuous Hermite polynomials}
Continuous q-Hermite are defined as:
$$
x H_n^{q}(\infty) = H_{n+1}^{q}(x) + \frac{q^n-1}{q-1} H_{n-1}^{q}(x), H_0=1, H_1=x
$$
$$
[n]_q! \delta_{n,m}= \int_{-\frac{!}{\sqrt{1-q}}}^{\frac{!}{\sqrt{1-q}}} H_n^{(q)}(x) H_m^{(q)}(x) d \mu_q^{c} (x),
$$
where 
$$
d \mu_q^{c} (x) = \frac{1}{2\pi} q^{-\frac{1}{8}} \theta_1 (\frac{\theta}{\pi},\frac{1}{2\pi i} \log q) dx = 
$$
$$
= \frac{1}{\pi}\sqrt{1-q}\sin(\theta) \prod_{n=1}^\infty (1-q^n)| 1-q^n \exp(2\pi\theta)|^2 dx
$$
for $0 \leq q < 1$, $\theta_1$ -- Jacobi theta one function.\\ $2\cos v = x \sqrt{1-q}$,\;\; $\supp \mu_q^c = [-\frac{2}{\sqrt{1-q}},\frac{2}{\sqrt{1-q}}]$ (see \cite{AAR} and \cite{BKS} for more details). 

\begin{theorem}[\cite{BY}]
    If $-1 \leq q \leq 1$, $s>0$, then there exist operators $A^\pm(f)=A^\pm_{q,s}(f)$, $g,f \in \mathbb{R}^N$, $N=\infty,1,2,\ldots$:
    
    $$A(f)A^+(g)-(s q)A^+(g)A(f)=s^N<f,g>I.$$

    $$A(f) \Omega = 0$$
    where $\mathcal{H}^{\otimes 0} = \mathbb{C}\Omega$.
\end{theorem}

\begin{construction}
    Take $q$-CCR operators: $a^\pm_q(f)=a(f)$.
    $$
    a(f) a^+(g) - q a^+(g) a(f) = <f,g> I, 
    $$
    on $\mathcal{F}_q(\mathcal{H})=\bigoplus_{n=0}^{\infty} \mathcal{H}^{\otimes n}$, as in Bożejko-Speicher \cite{BS} with scalar product
    $$
    <f_1 \otimes \ldots \otimes f_n | g_1 \otimes \ldots \otimes g_n>_q = <P_q^{(n)}(f_1 \otimes \ldots \otimes f_n) | g_1 \otimes \ldots \otimes g_n>.
    $$
    where $P_q^{(n)}=\sum_{\sigma \in S(n)} q^{inv(\sigma)} \sigma$. Here inverse of permutation $\sigma$ is defined as 
    $inv(\sigma)=\#\{\pi \in S(n): i<j\;{\rm and}\; \pi(i)>\pi(j)\}$ and $S(n)$ is a permutation group om $n$ letters. Now we define annihilation operator $A^\pm(f)=A^\pm_{q,s}(f)$ as
$$
A^\pm_{q,s}(f)=s^{N-1} a^\pm_q(f),\; s>0
$$
where $N$ on $\mathcal{H}^{\otimes n}$ is defined as:
$$
N(x_1 \otimes \ldots \otimes x_n)=n(x_1\otimes \ldots \otimes x_n)
$$
where $a^\pm_q(f) = f \otimes \xi$ (this is $q$-creation) and $a_q(f) = [a^+_q(f)]^*$, $f \in \mathcal{H}_\mathbb{R}$ (this is $q$-annihilation). This conjugation for vectors $\xi,\eta \in \mathcal{H}^{\otimes n}$ is defined in the new scalar product $<\xi|\eta>_{q,s} = s^{\binom{n}{2}} <\xi|\eta>_q$.
\end{construction}

\section{Combinatorial results on $P_2(2n)$}

\begin{definition}[q-conditions cummulants - Ph.Biane, M.Anshelevich]
    If $\mu$ -- probability measure on $\mathbb{R}$ with all moments, then the $q$-continuous cummulants are defined as follows:
    $$
    \mu \rightarrow \left(R^{(q)}_\mu(n)\right)_{n=1}^\infty
    $$ 
    in such a way that:
    \begin{equation}\label{eq1}
    \int_{-\infty}^\infty x^n d \mu(x)=\sum_{\mathcal{V} \in P(n)} q^{cr(V)} R_\mu^{(q)}(\mathcal{V}),
    \end{equation}
    where $P(n)$ is the set of all set-partitions on $\{1,2,\ldots,n\}$, and 
    $$
    R_\mu^{(q)}(\mathcal{V})=\prod_{B \in \mathcal{V}} R_\mu^{(q)}(|B|),
    $$
    where $\mathcal{V}$ -- partition of $\{1,2,\ldots,n\}$, and $cr(\mathcal{V})$ is a number of of hyperbolic (restricted) crossings defined by Ph. Biane \cite{PhB}.
\end{definition}
\begin{figure}[h!]
    \centering
    \includegraphics[width=0.7\textwidth]{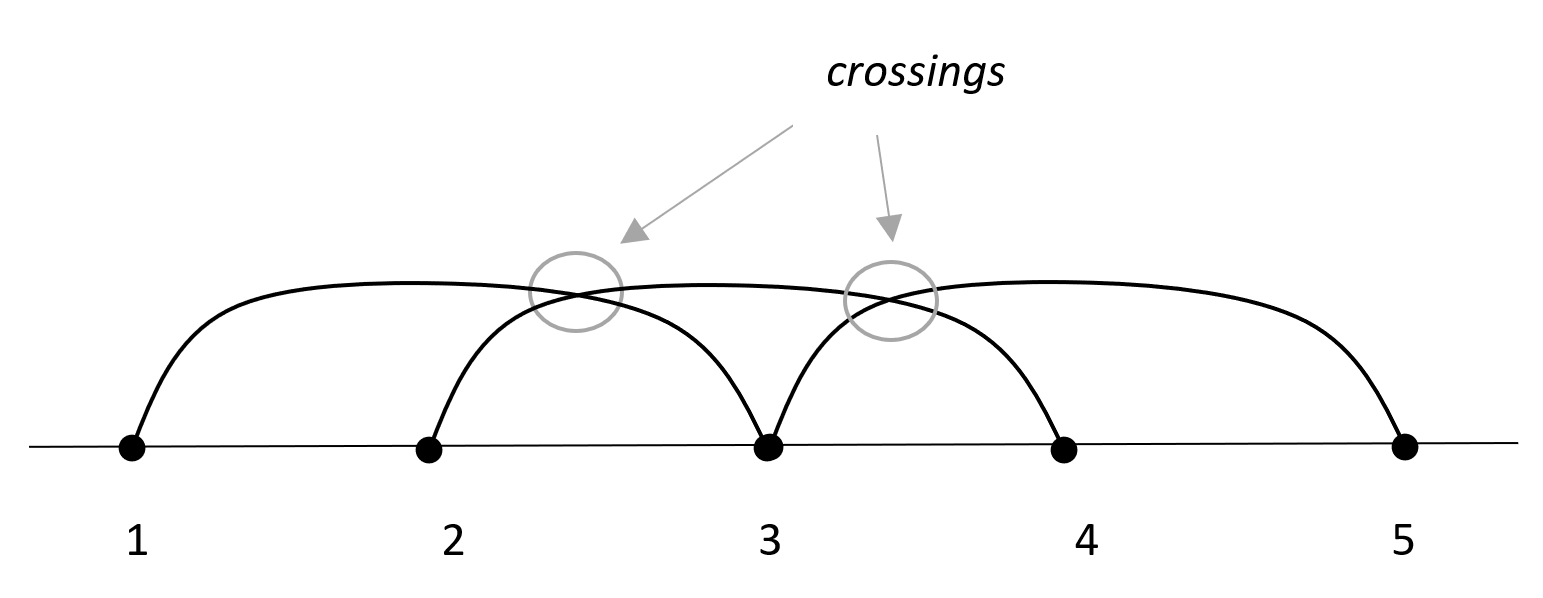}
    \caption{Crossings}
    \label{fig1}
\end{figure}
\begin{remark}
    A.Nica defined left-reduced number of crossing $c_0(\mathcal{V})$ as:
    $
        c_0(\mathcal{V}) = \#\{(m_1,m_2,m_3,m_4):1 \leq m_1 \leq m_2 \leq m_3 \leq m_4 \leq n: 
    $
    $(m_1,m_3) \in \mathcal{V}, (m_2,m_4) \in \mathcal{V}, (m_2,m_3) \notin \mathcal{V}, {\rm \;each\;} m_1, m_2 {\rm \;minimal\;in\;the} 
    $
    $
    {\rm \;class\;of\;}  \mathcal{V} {\rm \;containing\;it}\},
    $ 
    then Nica's q-cummulants $\widetilde{R}_\mu^{(q)}(n)$  come from (\ref{eq1}), where $cr(\mathcal{V})$ is replaced by $c_0(\mathcal{V})$.
\end{remark}
F.Oravecz \cite{O2} showed that Nica's q-cummulants \emph{are not} positivity preserving,~i.e.

If we define a ,,q-convolututon'': $\mu=\mu_1 *_q \mu_2$:  (Ph. Biane idea) is done as:
\begin{equation}\label{eq2}
   R^{(q)}_{\mu_1}(n) + R^{(q)}_{\mu_2}(n)  = R^{(q)}_{\mu}(n),\;\; n=1,2,3,\ldots, 
\end{equation}
then we have the following open problem:
\begin{problem}[open]
    Is Ph. Biane ,,$ *_q$-convolutions'' positivity preserving?
\end{problem}

Recently Carnovole and Koornwinder defined $q$-Discrete version of (\ref{eq2}):\\
$\mu=\mu_1 *_q^{disc} \mu_2$,\;\; $d \mu_i(x)=f_i(x) dx$,\;\; $f_1 *_q^{disc} f_2 = f$.
\begin{equation}\label{eq3}
    m_n^{disc}(f)=q^{\binom{n}{2}} \int_{-\infty}^\infty x^n f(x) d_q(x) = \sum_{k=0}^n 
    \genfrac{[}{]}{0pt}{}{n}{k}_q m_k^{disc} (f_1) m_{n-k}^{disc} (f_2),
\end{equation}
where $d_q(x)$ is the Jackson integral.

\noindent For $q=1$ we have classical convolutions.

\noindent For $q=0$ we have Boolean convolutions.

Now, we are describing the new $q$-convolution corresponding to $q$-Discrete Hermite polynomials of the type I. We give also Wick formula for that case.

\begin{theorem}[Bo\.{z}ejko--Yoshida\cite{BY} (Wick formula)]
    If $G(f)=A(f)+A^+(f)$, then
    $$
    <G(f_1)\ldots G(f_{2n}) \Omega | \Omega>=\sum_{\mathcal{V} \in P_2(2n)} s^{\frac{1}{2}ip(\mathcal{V}} \cdot q^{cr(\mathcal{V})} \prod_{(i,j) \in \mathcal{V}} <f_i | f_j>
    $$
    where $ip(\mathcal{V})=\sum_{(i,j) \in \mathcal{V}} inpt(i,j)$, $inpt(i,j) = \#\{$of k with $i<k<j\}$
    $$
    =\sum_{k=1}^n (j_k - i_k -1),\;\; {\rm if}\; \mathcal{V}=\{(i_1,j_1),\ldots,(i_n,j_n)\} \in P_2(2n).
    $$
\end{theorem}
For the proof see \cite{BY}.

We are recalling the crossing number definition for 2-partitions $\mathcal{V}$ (see \cite{BB}):
$$
cr(\mathcal{V})=\#\{(a,b),(c,d) \in \mathcal{V}: a < c < b < d,\includegraphics[width=3cm]{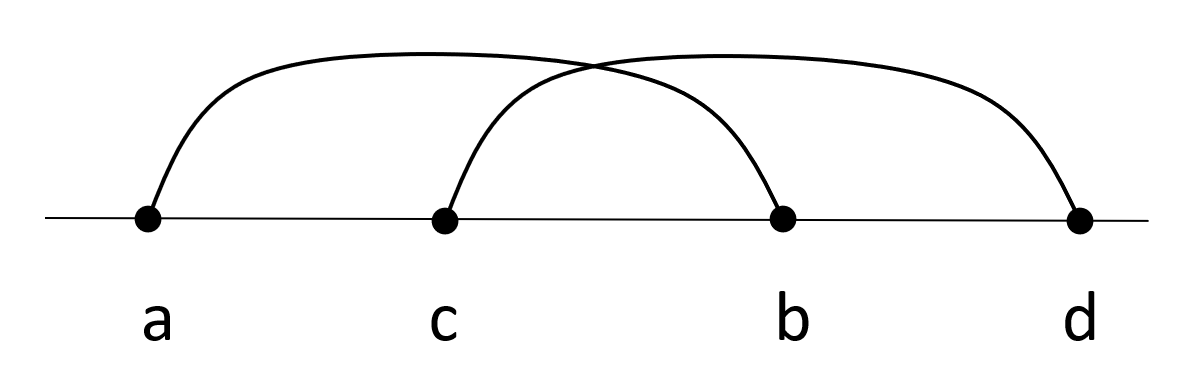}\}.
$$
\begin{theorem}[\cite{Bozejko2007}]
    If $\mathcal{V} \in P_2(2n)$, then
    $$
    cr(\mathcal{V}) + pbr(\mathcal{V})=\frac{1}{2}ip(\mathcal{V})    
    $$
    where $pbr(\mathcal{V})=\#\{(a,b),(c,d) \in \mathcal{V}: a<c<d<b, \includegraphics[width=3cm]{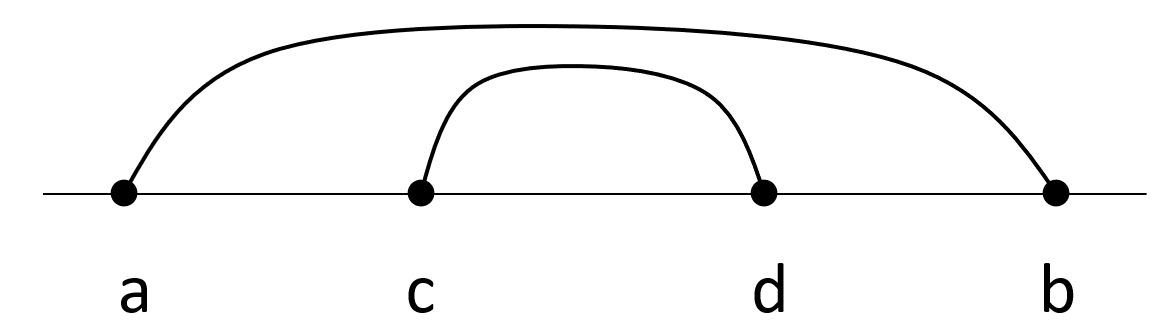}\}$.
\end{theorem}
The $pbr(\mathcal{V})$, also denoted as $nest(\mathcal{V})$, was introduced by A.Nica \cite{N}, de Medicist and Viennot \cite{deMV}. For more detailes on nesting, see papers: N.Blitric \cite{Bl}, Bożejko and Ejsmont \cite{BEj}. For the proof, we need the following Lemma, more details, see Bożejko~\cite{Bozejko2007}.

\begin{lemma}
    If
    $$
     \sum_{\mathcal{V} \in P_2(2n)} t_1(\mathcal{V}) \prod_{(i,j) \in \mathcal{V}} <f_i | f_j>
     = \sum_{\mathcal{V} \in P_2(2n)} t_2(\mathcal{V}) \prod_{(i,j) \in \mathcal{V}} <f_i | f_j>,
    $$
    for all vectors $f_i \in $Hilbert space, than
    $$
    t_1(\mathcal{V}) = t_2(\mathcal{V})
    $$
    for all 2-partitions $\mathcal{V} \in P_2(2n)$.
\end{lemma}

\noindent From this general construction we obtain $q$-CCR relation for $|q| \geq 1$.

\begin{theorem}[\cite{Bozejko2007}]
    If $q \geq 1$, then there exist operators on a proper Fock space satisfying the ($q$-CCR):
    $$
    B(f) B^+(g) - B^+(g) B(f) = q^N <f,g> I,\;\; f,g, \in \mathcal{H}\;{\rm (Hilbert~space),}
    $$
where $N(x_1 \otimes \ldots \otimes x_n) = n(x_1 \otimes \ldots \otimes x_n)$.

Moreover $\widetilde{G}(f) = B(f) + B^+(f)$:
$$
    <\widetilde{G}(f_1) \ldots \widetilde{G}(f_{2n}) \Omega | \Omega> = \sum_{\mathcal{V} \in P_2(2n)} q^{pbr(\mathcal{V})} \prod_{(i,j) \in \mathcal{V} <f_i | f_j>}.
$$
   
\end{theorem}

\noindent Proof's idea: Consider $A_{1/q,q}(f)$, submitting $1/q$ instead of $q$, $s=q$, where $B^\pm(f)=A^\pm_{1/q,q}(f)$, $f \in \mathcal{H}$ were constructed in Theorem~1.

\section{$q$-Discrete Hermite polynomials, $0 \leq q \leq 1$}\label{sect3}

We recall the definition of $q$-Discrete Hermite polynomial of type I and type II for $0 \leq q \leq 1$ as:

\noindent {\bf I type:} $h_0=1$, $h_1(x)=x$,
$x h_n(x) = h_{n+1}(x) + q^{n-1}[n]_q h_{n-1}(x)$,\;\;\\ $[n]_q=\frac{q^n-1}{q-1}=1+q+\ldots+q^{n-1}$.
Later we will denote ${h_n(x;q)}= h_n(x)$.\\

\noindent {\bf II type:} $\widetilde{h}_n(x;q)=i^{-n} h_n(ix;q^{-1})$, where $i=\sqrt{-1}$.\\

Now we recall the definition of two exponential functions 
$$
e_q(x) = \sum_{k=0}^\infty \frac{x^k}{(q:q)_k},\;\; E_q(x) = \sum_{k=0}^\infty \frac{q^{\binom{n}{2}} x^k}{(q:q)_k}
$$
where $(a:q)_k = (1-a)(1-aq)\ldots (1-aq^{k-1})$.

$$
[k]_q!=\frac{(q:q)_k}{(1-q)^k},\;\; \genfrac{[}{]}{0pt}{}{n}{j}_q = \frac{[n]_q!}{[j]_q! [n-j]_q!}\;\; {\rm (Gauss~symbol)}.
$$
\noindent{\bf Facts (see Andrews et al. \cite{AAR}):}
\begin{enumerate}
    \item $E_q(z)=\prod_{n=0}^\infty (1+q^n z)$, $z \in \mathbb{C}$,
    \item $e_q(z)E_{q}(-z)=1$, $z \in \mathbb{C}$,
    \item (I type) $\int_{-1}^1 h_m(x:q)h_n(x:q) E_{q^2}(-q^2x^2)d_qx = b_q \cdot q^{\binom{n}{2}}(q:q)_n \delta_{n,m}$,
    \item (II type) $\int_{-\infty}^\infty \widetilde{h}_m(x:q) \widetilde{h}_n(x:q) e_{q^2} (-x^2) d_q x =
    c_q \cdot q^{-n^2}(q:q)_n \delta_{n,m}$,
    where
    $$\int_0^x f(x) d_q(x) = (1-q) \sum_{k=0}^\infty f(q^k x) q^k x$$ 
    is well known Jackson integral for functions with support $\supp(f) \subset \mathbb{R}^+$,
    and for arbitrary $f:\mathbb{R} \rightarrow \mathbb{C}$ we define
    $$\int_{-\infty}^\infty f(x) d_q(x) = (1-q) \sum_{k=-\infty}^\infty \sum_{\varepsilon = \pm 1} q^k f(\varepsilon q^k); \;\; \supp(f) \subset \mathbb{R}.$$
\end{enumerate}

\noindent\textbf{Commutation relation to the Fock representation of type I discrete Hermite polynomials.}\\

In Theoreom 1 put $s=q$, $q=q$, $0 \leq 1 \leq 1$, then operators
$$
C_q^\pm(f)=A_{q,q}^\pm(f),\;\; \widehat{G}(f)=C(f)+C^+(f)
$$
satisfy the following theorem:
\begin{theorem}
    \begin{enumerate}
        \item If $||f_i||=1$, $i=1,2,\ldots,2n$, then
        $$
        <\widehat{G}(f_1)\ldots \widehat{G}(f_{2n}) \Omega| \Omega>=\sum_{\mathcal{V} \in P_2(2n)} q^{\frac{1}{2} ip(\mathcal{V})+cr(\mathcal{V})} \prod_{(i,j) \in \mathcal{V}} <f_i | f_j>
        $$
        \item 
        $$
        \int x^{2n} d \mu_q^I(x)=<\widehat{G}(f_1)^{2n}  \Omega| \Omega>=[1]_q [3]_q\ldots [2n-1]_q =
        \sum_{\mathcal{V} \in P_2(2n)} q^{e_0}(\mathcal{V}),
        $$
        where $e_0(\mathcal{V})$ was introduced by \cite{deMV}, where 
        $$e_0(\mathcal{V})=pbr(\mathcal{V})+
        2cr(\mathcal{V})=\frac{1}{2}ip(\mathcal{V})+cr(\mathcal{V}).
        $$
        \item Moreover
        $$
        A_q(f) A_q^+(g) - q^2 A_q^+(g) A_q(f) = q^N <f,g>I.
        $$
        for $f,g \in \mathcal{H}$.
    \end{enumerate}
\end{theorem}

\noindent \textbf{Problems:}
\begin{enumerate}
    \item Prove positivity of $q$-Discrete (continuous) convolutions for $0 < q < 1$?
    \item Describe $q$-Discrete Poisson measure (process)?
    \item Calculate the operator norm of $||\widehat{G}(f_i)||=?$, $i=1,2,\ldots$
    \item If $\Omega$ is faithful state in the corresponding Fock space?
\end{enumerate}

\section{Another q-analogues of classical convolutions}

Let us define Jackson ,,$q$-moments'' for ,,good'' function $f:\mathbb{R}\rightarrow \mathbb{R}$ as follows:
$$
m_n^{disc}(f)=q^{\binom{n}{2}} \int_{-\infty}^\infty f(x) x^n d_q(x),
$$
and ,,$q$-Discrete convolutions'' of Carnowale and Koornwinder \cite{CK}
$$
(f \otimes_q g)(x) = \sum_{n=0}^\infty \frac{(-1)^n m_n^{disc}(f)}{[n]_q!} (\delta_q^n g) (x)
$$
where
$$
\delta_q f(x) = \left \{
\begin{array}{ll}
\frac{f(x)-f(qx)}{x-qx},& x \neq 0, \lim_{q\rightarrow 1} \delta_q f(x) = f'(x),\\
f'(0),& x=0.
\end{array}
\right.
$$
Note that if $q=1$, we have
$$
\left( \int_{-\infty}^\infty dt f(t) \frac{(-1)^n t^n}{n!} \right) g^{(n)}(x)=
\int_{-\infty}^\infty dt f(t) \left( \sum_{n=0}^\infty \frac{(-t)^n}{n!} g^{(n)}(x) \right)=
$$
$$
=\int_{-\infty}^\infty dt f(t)\; g(x-t) = (f * g)(x).
$$
which is the classical convolution.

\begin{theorem}[Carnovale, Koornwinder \cite{CK}] For ,,good'' functions $f,g: \mathbb{R} \rightarrow \mathbb{R}$ 
    $q$-Discrete convolution is \textbf{associative} and \textbf{commutative}. Moreover
    $$
    m_n^{disc} (f \otimes_q g) = \sum_{n=0}^n \genfrac{[}{]}{0pt}{}{n}{j}_q m_k^{disc}(f)\; m_{n-k}^{disc}(g).
    $$
\end{theorem}
If $q=0$, we get Boolean convolution. 

\noindent If $q=1$, we get classical convolution on $\mathbb{R}$.\\

\noindent \textbf{Problem:}
    
    Find characterization $q$-Discrete moments sequence $m_n^{disc}(f)$, i.e. for $f \geq 0$
    $$
    m_n^{disc}(f) = \int_{-\infty}^\infty f(x) x^n d_q(x) ?
    $$

\section{Braided Hopf algebras of Kempf and Majid}

\begin{definition}
    Braided line is a braided algebra $\mathcal{A}=\mathbb{C}[[x]]$ formal power series in variable $x$
    which has braiding
    $$
    \Phi (x^k \otimes x^l) = q^{kl} x^l \otimes x^k,
    $$
    commultiplation:
    $$
    \Delta(x^k)=\sum_{j=0}^k \genfrac{[}{]}{0pt}{}{k}{j}_q x^{k-j} \otimes x^j
    $$
    co-unit
    $$
    \varepsilon(x^k)=\delta_{k,0}
    $$
    braided antipode
    $$
    S(x^k)=(-1)^k q^{\binom{k}{2}} x^k = (-1)^k q^{\frac{k(k-1)}{2}} x^k,
    $$
    and then we get  the $q$-analogue of of Taylor's formula:
    $$
    \Delta(f(x))=\sum_{j=0}^\infty \frac{x^j}{[j]_q!} \otimes \delta_q(f(x)).
    $$
\end{definition}

\begin{theorem}[Kemp, Majid \cite{KM} ]
    If $Q f(x) = f(qx)$, then we have
    $$
        (f*_q g)(x)= (f \otimes id)(m \otimes id)[id \otimes Q \otimes id](id \otimes S \otimes id)(id \otimes \Delta)(f \otimes g)(x)
    $$
\end{theorem}
Moreover as observed by Koornwinder we have
$$
\Delta(e_q(x))=e_q(x) \otimes e_q(x),
$$
$$
S(e_q(x))=E_q(-x),
$$
$$
\varepsilon(e_q(x))=1.
$$

\section{The construction of $q$-Discrete Fock space\linebreak $\mathcal{F}_q^{disc}(\mathcal{H})$ and $q$-Discrete Brownian motions}

Now we present for $0 \leq q \leq 1$ the construction of $q$-Discrete Fock space $\mathcal{F}_q^{disc}(\mathcal{H})$ for $q$-Discrete Hermite of Type I, which is the completion of the full Fock space\linebreak $\mathcal{F}(\mathcal{H})=\bigoplus_{n=0}^\infty \mathcal{H}^{\oplus n}=\mathbb{C} \Omega \oplus \mathcal{H} \oplus (\mathcal{H} \otimes \mathcal{H}) \ldots$ under the positive inner product on $\mathcal{H}^{\oplus n}$ done by:
$$
<x_1 \otimes \ldots \otimes x_n | y_1 \otimes \ldots \otimes y_m>_q = 
$$
$$
=\delta_{n,m} q^{\binom{n}{2}} \sum_{\pi \in S(n)} q^{inv(\pi)} <x_1|y_{\pi(1)}>\cdot \ldots \cdot <x_n|y_{\pi(n)}>. 
$$
We define creation operator $A_q^+(f)\xi_n=f \otimes \xi_n$, $f \in \mathcal{H}$, $\xi_n \in \mathcal{H}^{\otimes n}$
and the annihilation operator 
$$
A_q(f)x_1 \otimes \ldots \otimes x_n= q^{n-1} \sum_{k=1}^n q^{k-1} <x_k | f> x_1 \otimes \ldots \otimes \check{x}_k \otimes \ldots \otimes x_n
$$
In the paper [B-Y] we have more general construction
$$
\widetilde{A}_{q,s}(f)x_1 \otimes \ldots \otimes x_n= s^{2(n-1)} \sum_{k=1}^n q^{k-1} <x_k | f> x_1 \otimes \ldots \otimes \check{x}_k \otimes \ldots \otimes x_n
$$
If we put $s^2=q$ we get our $q$-Discrete Fock space $\mathcal{F}_q^{disc}(\mathcal{H})$.
\begin{remark}
    For $f,g \in \mathcal{H}$ we have the following $q$-Discrete Commutation Relation:
    $$
      A(f)A^+(g)-q^2 A^+(g)A(f)=q^N<f,g>I.
    $$
\end{remark}
See more details in \cite{BY, Bl} for a more general case.

Moreover, we have $q$-Discrete Gaussian random variables $\widehat{G}_q(f)=A_q(f)+A_q^+(f)$. We get $q$-version of Wick formula
$$
    <\widehat{G}_q(f_1)\ldots \widehat{G}_q(f_{2n}) \Omega | \Omega>=\sum_{\mathcal{V} \in P_2(2n)} q^{\frac{1}{2}ip(\mathcal{V})} \cdot q^{cr(\mathcal{V})} \prod_{(i,j) \in \mathcal{V}} <f_i | f_j>.
$$
Our Gaussian $\widehat{G}_q(f)$ at the vacuum state $\Omega$ has  the spectral measure $\mu^{disc}_q$ corresponding to $q$-Discrete Hermite polynomials of type I which were defined by following recurrence:
$$x h_n(x) = h_{n+1}(x) + q^{n-1}[n]_q h_{n-1}(x),\;\;[n]_q=\frac{q^n-1}{q-1}=1+q+\ldots+q^{n-1},
$$
$h_0(x)=1$, $h_1(x)=x$, in the Section~\ref{sect3}. 

Now we define {\bf $q$-Discrete Brownian motion} $BM_t$ as follows.\\ Take
$
\mathcal{H}=L^2(\mathbb{R}^+,dx)
$
and $f=\chi_{[0,t)}$,
$$
f(x)=\left\{
\begin{array}{ll}
1&{\rm for}\; x \in [0,t),\\
0&{\rm otherwise.}
\end{array}\right.
$$
Then
$$
BM_t=\widehat{G}_q(\chi_{[0,t)})
$$
is our $q$-Discrete Brownian motion.
\begin{remark}
Case $q=1$ is the {\bf classical} Brownian motion, and $q=0$ is the {\bf Boolean} Brownian motion.
\end{remark}

\noindent {\bf Problem:} Is the von Neumann algebra $BB_q =$ WO-closure of $\{BM_t: t \geq 0\}$ is factorial, that means it has one-dimensional center?

This $BB_q$ algebra corresponds to $q$-Discrete Hermite polynomials of type I, but the corresponding problem for continuous $q$-Discrete Hermite polynomials was solved by Bożejko, Kümmerer and Speicher \cite{BKS}.

\section{Matrix version of Khintchine inequalities}

We are looking for matricial version Khintchine inequalities for random variables $X_1,X_2,\ldots,X_n$,
\begin{equation}\label{eq_1}
\left\|\sum_{j=1}^n \alpha_j \otimes X_j\right\| \cong \max\left\{\left\|\sum_{j=1}^n \alpha_j \alpha_j^*\right\|^{\frac{1}{2}},\left\|\sum_{j=1}^n \alpha_j^* \alpha_j\right\|^{\frac{1}{2}}\right\}
\end{equation}
for $n=1,2,\ldots$ and $\alpha_j$ are complex matrices of arbitrary sizes and the norms are operator norms, where
$a \cong b$ iff $K_1b \leq a \leq K_2 b$ for some $K_1,K_2 > 0$.

\begin{theorem}
Inequality (\ref{eq_1}) holds for $q$-continuous, $q$-discrete Gaussian, Kesten Gaussian, Booleand Gaussian
and many others examples.
\end{theorem}

We  give only proof for Booleand Gaussian ($0=q$-discrete Gaussian) which is much more, since we have isometrical-isomorphism, that is  for $\|f_i\|=1$, $f_i \in \mathcal{H}$, $a_i = a(f_i)$, $G^B(f_i) = a_i + a^{+}_i$
\begin{equation}
\left\| \sum_{i=1}^N \alpha_i \otimes G^B(f_i)\right\|
= \max\left\{\left\|\sum_{j=1}^n \alpha_j \alpha_j^*\right\|^{\frac{1}{2}},\left\|\sum_{j=1}^n \alpha_j^* \alpha_j\right\|^{\frac{1}{2}}\right\}.
\end{equation}

\noindent{\bf Proof.} Our Boolean Fock space $\mathcal{F}_0(\mathcal{H}=\mathbb{C}\Omega \oplus \mathcal{H}=\mathbb{C}\Omega\oplus lin\{e_1,e_2,\ldots,e_n\}$. 
Let $G^B(f_i)(\Omega)=\omega_i(\Omega)=(a_i+a^+_i)(\Omega)=e_i$. Since by definition $a_i(\Omega)=0$ and $a^+_i(\Omega)=e_i$ so
$$
\omega_i(e_k)=a_i(e_k)+a^+_i(e_k)=\delta_{ik}\Omega.
$$
Then
$$
\omega_i^2(\Omega)=\omega_i(e_i)=\Omega.
$$
Therefore
$$
\omega_i^2(e_k)=\omega_i(\omega_i(e_k))=\omega_i(\delta_{ik}\Omega)=\delta_{ik} e_i = e_k.
$$
So
$$
\omega_i^2=\mathbb{I}.
$$
From that consideration we get
$$
\omega_1=\left(
\begin{array}{c|c|c}
     0&1&0 \\ \hline
     1&0&0 \\  \hline
     0&0&0
\end{array}~~~~\right),
$$
$$
\omega_2=\left(
\begin{array}{c|c|c}
     0&0&1 \\ \hline
     0&0&0 \\  \hline
     1&0&0
\end{array}~~~~\right).
$$
Let 
$$
T=\sum_{i=1}^N \alpha_i \otimes\omega_i=\left(
\begin{array}{c|c|c|c|c}
     0&\alpha_1&\alpha_2&\alpha_3&\ldots\alpha_N \\ \hline
     \alpha_1&0&0&0&0 \\  \hline
     \alpha_2&0&0&0&0 \\  \hline
     \vdots&&&& \\  \hline
     \alpha_N&0&0&0&0 \\  
\end{array}~~~~\right),
$$
$$
T^*=\left(
\begin{array}{c|c|c|c|c}
     0&\alpha_1^*&\alpha_2^*&\alpha_3^*&\ldots\alpha_N^* \\ \hline
     \alpha_1^*&0&0&0&0 \\  \hline
     \alpha_2^*&0&0&0&0 \\  \hline
     \vdots&&&& \\  \hline
     \alpha_N^*&0&0&0&0 \\  
\end{array}~~~~\right).
$$
So we get
$$
T T^*=\left(
\begin{array}{c|c|c|c|c}
     \sum_{i=1}^N \alpha_i \alpha_i^*&0&0&\ldots&0 \\ \hline
     0&\alpha_1 \alpha_1^*&\alpha_1 \alpha_2^*&\ldots&\alpha_1 \alpha_N^* \\  \hline
     0&\alpha_2 \alpha_1^*&\alpha_2 \alpha_2^*&\ldots&\alpha_2 \alpha_N^* \\  \hline
     \vdots&&&& \\  \hline
     0&\alpha_N \alpha_1^*&\alpha_N \alpha_2^*&\ldots&\alpha_N \alpha_N^*  
\end{array}~~~~\right)=\left(
\begin{array}{c|c}
     A&0  \\ \hline
     0&B 
\end{array}\right),
$$
$A=(\sum_{i=1}^N \alpha_i \alpha_i^*)$, $B=(\alpha_i \alpha_j^*)_{i,j=1,2,\ldots,N}$.\\

From this form we obtain
$\|T T^* \| = \max\{\|A\|,\|B\|\}$. It is easy to see that $\|A\|=\|\sum_{i=1}^N \alpha_i \alpha_i^*\|$. To calculate $\|B\|$ we observe, that $\|(\alpha_i \alpha_j^*)\|=$
$$
= \left\| \left(
\begin{array}{cccc}
     \alpha_1& 0 & \ldots & 0 \\
     \alpha_2& 0 & \ldots & 0\\
     \ldots & 0 & \ldots & 0\\
     \alpha_N & 0 & \ldots & 0\\
\end{array}\right)
\left(
\begin{array}{cccc}
     \alpha_1& 0 & \ldots & 0 \\
     \alpha_2& 0 & \ldots & 0\\
     \ldots & 0 & \ldots & 0\\
     \alpha_N & 0 & \ldots & 0\\
\end{array}\right)^*
\right\|=
$$
$$
= \left\| \left(
\begin{array}{cccc}
     \alpha_1^*& \alpha_2^* & \ldots & \alpha_N^* \\
     0 & 0 & \ldots & 0\\
     \ldots & 0 & \ldots & 0\\
     0 & 0 & \ldots & 0\\
\end{array}\right)
\left(
\begin{array}{cccc}
     \alpha_1& 0 & \ldots & 0 \\
     \alpha_2& 0 & \ldots & 0\\
     \ldots & 0 & \ldots & 0\\
     \alpha_N & 0 & \ldots & 0\\
\end{array}\right)
\right\|=
$$
$$
=\left\|\sum_{i=1}^N \alpha_i^* \alpha_i \right\| = \|B\|
$$
so we get our theorem for Boolean Gaussian. For other cases proofs are a little bit more complicated, see \cite{BKW,Bozejko2012}.

\begin{cor}
    If $VN(X_1,X_2,\ldots,X_n)$ has trace, then for some $q=q(N)$,
    $VN(X_1,X_2,$ $\ldots,X_n)$ is NOT injective and also it is  a FACTOR.
\end{cor}

\section*{Statements and Declarations}

The work was not financed from any fund. We declare no conflict of interest. No data was generated for research purposes.

%All authors contributed to the study conception and design. Material preparation, data collection and analysis were performed by [full name], [full name] and [full name]. The first draft of the manuscript was written by [full name] and all authors commented on previous versions of the manuscript. All authors read and approved the final manuscript.


\begin{thebibliography}{40}
\bibitem[An]{Anshelevich}
Anshelevich, Michael. Partition-dependent stochastic measures and -deformed cumulants. Documenta Mathematica 6 (2001): 343--384.

\bibitem[AB]{Accardi}
Accardi, Luigi,  Bożejko, Marek.
Interacting Fock spaces and Gaussianization of probability measures. Infin. Dimens. Anal. Quantum Probab. Relat. Top.1(1998), no.4, 663--670.

\bibitem[AAR]{AAR}
Andrews, G., Askey, R., Roy, R. (1999). Special Functions (Encyclopedia of Mathematics and its Applications). Cambridge: Cambridge University Press.

\bibitem[B1]{Bozejko2001}
Bożejko, Marek.
Deformed free probability of Voiculescu. Infinite dimensional analysis and quantum probability theory.  Sūrikaisekikenkyūsho Kōkyūroku (2001), no.1227, 96–113.

\bibitem[B2]{Bozejko2007}
Bożejko, Marek. Remarks on $q$-CCR relations for $|q| > 1$. Banach Center Publications 78,1 (2007) 59--67.

\bibitem[B3]{Bozejko2012}
Bożejko, Marek. Deformed Fock spaces, Hecke operators and monotone Fock space of Muraki. Demonstratio Mathematica, vol. 45, no. 2, 2012, pp. 399--413.

\bibitem[BB]{BB}
Bożejko M., Bożejko W., Generalized Gaussian processes and relations with random matrices and positive definite functions on permutation groups, Infinite Dimensional Analysis, Quantum Probability and Related Topics Vol. 18, No. 3 (2015), 1550020-1 -- 1550020-19.

\bibitem[Bl]{Bl}
Blitvic N. The (q,t)-Gaussian Process. Journal of Functional Analysis. 2012 Nov 15;263(10):3270-3305.

\bibitem[BKS]{BKS}
Bożejko, M., Kümmerer, B., and Speicher, R., q-Gaussian processes: non-commutative and classical aspects, Comm. Math. Phys., 185(1), (1997)

\bibitem[BY]{BY}
Bożejko, Marek, Yoshida, Hiroaki. Banach Center Publications 73 (2006), 127-140

\bibitem[BEj]{BEj}
Bożejko, Marek, Ejsmont, Wiktor.
The double Fock space of type B. SIGMA Symmetry Integrability Geom. Methods Appl.19(2023), Paper No. 040, 22 pp.

\bibitem[BKW]{BKW}
Bożejko, Marek, Krystek, Anna, Wojakowski, \L{}ukasz. Remarks on the r and $\Delta$ convolutions. Math. Z. 253, 177–196 (2006).

\bibitem[BS]{BS}
Bożejko, Marek, Speicher, Roland. An example of a generalized Brownian motion. Comm. Math. Phys. 137 (1991), no. 3, 519–531.

\bibitem[PhB]{PhB}
Biane, Philippe. Some properties of crossings and partitions. Discrete Math. 175 (1997), no. 1-3, 41–53.

\bibitem[CK]{CK}
Carnovale, G. Koornwinder, T. H. 
A q-analogue of convolution on the line. 
Methods Appl. Anal. 7 (2000), no. 4, 705–726.

\bibitem[KM]{KM}
Kempf, Achim, Majid, Shahn. Algebraic q-integration and Fourier theory on quantum and braided spaces. J. Math. Phys. 35 (1994), no. 12, 6802–6837

\bibitem[DP]{DP}
Díaz, Rafael, Pariguan, Eddy. On the Gaussian q-distribution. J. Math. Anal. Appl. 358 (2009), no. 1, 1–9.

\bibitem[deM,V]{deMV}
de Médicis, Anne, Viennot, Xavier G. 
Moments des q-polynômes de Laguerre et la bijection de Foata-Zeilberger. 
Adv. in Appl. Math. 15 (1994), no. 3, 262–304.

\bibitem[EP]{EP}
Effros, Edward G., Popa, Mihai.
Feynman diagrams and Wick products associated with q-Fock space. (English summary)
Proc. Natl. Acad. Sci. USA 100 (2003), no. 15, 8629–8633.

\bibitem[K1]{K1}
Kula, Anna. A limit theorem for the q-convolution. Noncommutative harmonic analysis with applications to probability III, 245–255, Banach Center Publ., 96, Polish Acad. Sci. Inst. Math., Warsaw, 2012.

\bibitem[K1]{K1}
Kula, Anna. The q-deformed convolutions: examples and applications to moment problem. Oper. Matrices 4 (2010), no. 4, 593–603.

\bibitem[N]{N}
Nica, Alexandru Crossings and embracings of set-partitions and q-analogues of the logarithm of the Fourier transform. Proceedings of the 6th Conference on Formal Power Series and Algebraic Combinatorics (New Brunswick, NJ, 1994). Discrete Math. 157 (1996), no. 1-3, 285–309.

\bibitem[O1]{O1}
Oravecz, Ferenc The number of pure convolutions arising from conditionally free convolution. Infin. Dimens. Anal. Quantum Probab. Relat. Top. 8 (2005), no. 3, 327–355.

\bibitem[O2]{O2}
Oravecz, Ferenc Nica's q-convolution is not positivity preserving. Comm. Math. Phys. 258 (2005), no. 2, 475–478.

\bibitem[Y]{Y}
Yoshida, Hiroaki Remarks on the s-free convolution. Non-commutativity, infinite-dimensionality and probability at the crossroads, 412–433, QP–PQ: Quantum Probab. White Noise Anal., 16, World Sci. Publ., River Edge, NJ, 2002. 
\end{thebibliography}
\end{document}